\documentclass[10pt,twocolumn,aps,prd,amsmath,amssymb,preprintnumbers,showpacs,nofootinbib]{revtex4-1}

\usepackage[utf8]{inputenc}
\usepackage[english]{babel}
\usepackage{amsmath}
\usepackage{amsfonts}
\usepackage{amssymb}
\usepackage{epsfig}
\usepackage{graphics,psfrag,rotating}
\usepackage{graphicx}
\usepackage{dcolumn}
\usepackage{bm}
\usepackage{epstopdf}
\usepackage{color}
\usepackage[usenames,dvipsnames,svgnames]{xcolor}
\usepackage[colorlinks=true,
            linkcolor=red,
            urlcolor=gray,
            citecolor=blue]{hyperref}
  \usepackage{hyperref}
\usepackage{graphics}
\usepackage{graphicx}
\usepackage{feynmf}
\usepackage{color}
\usepackage{subfigure}
\usepackage[utf8]{inputenc}
\usepackage[english]{babel}
\usepackage{amsmath}
\usepackage{amsfonts}
\usepackage{amssymb}
\usepackage{epsfig}
\usepackage{graphics,psfrag,rotating}
\usepackage{graphicx}
\usepackage{dcolumn}
\usepackage{bm}
\usepackage{epstopdf}
\newcommand{\be}{\begin{equation}}
\newcommand{\ee}{\end{equation}}
\newcommand{\ba}{\begin{eqnarray}}
\newcommand{\ea}{\end{eqnarray}}

\newcommand{\bw}{\begin{widetext}}
\newcommand{\ew}{\end{widetext}}

\newcommand{\Od}{{\mathcal O}}
\newcommand{\lsim}{\mbox{\raisebox{-.9ex}{~$\stackrel{\mbox{$<$}}{\sim}$~}}}

\begin{document}


\title{Vacuum energy as dark matter}


\author{F.\,D.\ Albareti$^{(a)}$}
\author{J.\,A.\,R.\ Cembranos$^{(b)}$}
\author{A.\,L.\ Maroto$^{(b)}$}

\affiliation{$^{(a)}$ Departamento de F\'isica Te\'orica, Universidad Aut\'onoma de Madrid, Campus de Cantoblanco, E-28049 Madrid, Spain}
\affiliation{$^{(b)}$Departamento de F\'{\i}sica
Te\'orica I, Universidad Complutense de Madrid, E-28040 Madrid,
Spain.}

\date{\today}


%
%

\begin{abstract}
We consider the vacuum energy of massive quantum fields in an expanding universe.
We define a conserved renormalized energy-momentum
tensor by means of a comoving cutoff regularization. Using exact solutions
for de Sitter space-time, we show that in a certain range of mass and renormalization scales there is a contribution to the vacuum energy density that scales
as nonrelativistic matter and that such a contribution becomes dominant at late times.
By means of the WKB approximation, we find that
 these results can be extended to arbitrary Robertson-Walker geometries.
We study the range of parameters in which the
vacuum energy density would be compatible with current limits on
dark matter abundance. Finally, by calculating the  vacuum energy in a perturbed Robertson-Walker
background, we obtain  the speed of sound of density perturbations and
show that the vacuum energy density contrast can grow on sub-Hubble scales as in
standard cold dark matter scenarios.

\end{abstract}

\maketitle



\section{Introduction}
\label{intro}


Since the discovery of the accelerated expansion of the Universe \cite{SNIa}, we have learned that most of the Universe content is a kind of cosmic fluid with negative pressure known as dark energy. Furthermore, even the dominant contribution to the  matter content is to a high degree unknown, being described by a weakly interacting  component which has received the name of dark matter. This present knowledge about the composition of the Universe has been possible thanks to precise measurements including for instance 
CMB temperature power spectrum \cite{Planck_WMAP7}, large-scale structures correlation functions (BAOs \cite{Eisenstein}) or high-redshift type Ia supernovae \cite{SNIa}. To the best knowledge of the authors a theoretical explanation of the values or even the presence of these dark components is absent.

Considering the simplest model of dark energy, i.e.\ a cosmological constant, it is believed that this term has classical and quantum contributions  \cite{Shapiro1,Shapiro2,Sola,Martin}. The classical one may just be taken as a parameter of the theory since the most general form of the General Relativity equations may include this contribution without breaking any of the fundamental assumptions of the
theory such as general covariance or energy-momentum conservation \cite{Lovelock,Wald,HE}. On the other hand, the quantum contribution is expected from quantum field theory grounds. However, as it is widely known, the theoretical predictions of its value and the measured one differ in many orders of magnitude. This difference may be compensated by the classical contribution leaving us with the observed value. The fine tuning necessary for this to happen is one of the drawbacks that have brought us to the cosmological constant problem.

However, many of the standard arguments about the  contribution of the  zero-point quantum
fluctuations to the cosmological constant are based on calculations performed in
flat space-time, in which the vacuum state
is assumed to respect the Lorentz invariance of the Minkowski space-time. In fact,
when taking into account
that the
actual geometry of the Universe is not Minkowskian and
moving to a Robertson-Walker background, new contributions to the vacuum
energy-momentum tensor appear \cite{Birrell} and
new aspects of the problem are revealed which were not apparent in the flat space-time calculations \cite{Maggiore,Hollenstein}.

One of the major problems in calculating the vacuum energy density is the divergent
integral over the Fourier modes appearing in the canonical quantization procedure.
Several methods have been proposed in the literature
in order to obtain finite renormalized results. Thus in general, the physical
renormalized vacuum expectation value of the energy-momentum tensor
$\langle 0\vert T_{\mu \nu}\vert 0\rangle_{\text{ren}}$
is obtained from the divergent bare quantities by subtracting the
regularized divergences by means of appropriate counterterms, i.e
\begin{eqnarray}
\langle 0\vert T_{\mu \nu}\vert 0\rangle_{\text{ren}}=\langle 0\vert T_{\mu \nu}\vert 0\rangle_{\text{bare}}+\langle 0\vert T_{\mu \nu}\vert 0\rangle_{\text{count}}\,.
\end{eqnarray}
Different schemes have been proposed to obtain the regularized bare quantities. For instance, in flat
space-time one of the simplest possibilities is to use a  cutoff on the three-momentum of the modes $\Lambda_P$.  However, it has been argued \cite{Akhmedov,Ossola} that the maximum value of the three-momentum is not a Lorentz invariant
quantity and therefore, the regularized bare contributions break the Lorentz symmetry of Minkowski
space-time. Indeed, in the case of a real minimally coupled scalar field,  the regularized bare quantities read
\begin{eqnarray}
\langle 0\vert T^{\mu}_{ \;\;\nu}\vert 0\rangle_{\text{bare}}=\mbox{diag}(\rho_{\text{bare}},-p_{\text{bare}},-p_{\text{bare}},-p_{\text{bare}})
\end{eqnarray}
with the leading contributions 
\begin{eqnarray}
\rho_{\text{bare}}=\frac{1}{16\pi^2}\left(\Lambda_P^4+m^2\Lambda_P^2-
m^4\ln\left(\frac{\Lambda_P}{\mu}\right)\right)
\end{eqnarray}
\begin{eqnarray}
p_{\text{bare}}=\frac{1}{16\pi^2}\left(\frac{\Lambda_P^4}{3}-\frac{m^2\Lambda_P^2}{3}+
m^4\ln\left(\frac{\Lambda_P}{\mu}\right)\right)
\end{eqnarray}
i.e.  in $\langle 0\vert T_{\mu\nu}\vert 0\rangle_{\text{bare}}$ only the logarithmic term
would be proportional
to $\eta_{\mu\nu}$.

This problem is avoided  in other regularization schemes, such as
dimensional regularization, which preserve the underlying symmetries of the
theory. Dimensional regularization has been carried out in flat space-time \cite{Akhmedov}
yielding a cosmological constant contribution with $p_{\text{bare}}=-\rho_{\text{bare}}$, where again
for scalar fields of \mbox{mass $m$}
\begin{eqnarray}
\rho_{\text{bare}}&=&-\frac{m^4}{64\pi^2}\left(\frac{2}{\epsilon}+\frac{3}{2}-\gamma-\ln\left(\frac{m^2}{4\pi\mu^2}\right)\right)
\end{eqnarray}
with $D=4-\epsilon$ the space-time dimension,  $\gamma$ the Euler-Mascheroni constant and $\mu$ the renormalization scale.
 In curved space-time, it can
be seen that the vacuum expectation value of the energy-momentum tensor
is no longer a simple cosmological constant term, but  in general is a nonlocal functional of the metric tensor. The calculation of the divergent local part
in dimensional regularization
shows that there is a contribution that behaves as a cosmological constant together with
other local and conserved tensors which depend on the curvatures.  Exact results
 including also the finite contributions have been obtained only for
conformally trivial systems \cite{Birrell}.

Notice, however, that strictly speaking the  problem we mentioned with the cutoff regularization would not be
present in a Robertson-Walker cosmological background, since
in this case Lorentz invariance is not a symmetry of the background metric. As a matter of fact,
in a Robertson-Walker background there is a special frame of reference (that at rest
with the CMB) which is most suitable for calculations. In this case, a three-dimensional momentum
cutoff defined over the homogeneous and isotropic spatial sections may have a more satisfying interpretation, since it respects the symmetries of the background geometry.

Accordingly, several recent works  have focused on the
possibility of using different kinds of cutoff regularizations in Robertson-Walker backgrounds.
Thus in \cite{Bilic1} a cutoff scale was used in the context of supersymmetric models,
and in \cite{Bilic2} a covariant cutoff scheme was proposed in general curved
space-times.
On the other hand, one may use a cutoff to perform the integration and
consider different renormalization prescriptions according to the counterterms included. Thus
for instance, in \cite{Maggiore}   the vacuum energy density obtained in a flat space-time is
subtracted in a similar process to the definition of the Arnowitt-Deser-Misner mass on asymptotically flat space-times.

Notice that in general, different renormalization schemes may provide different renormalized
expressions. On general grounds \cite{Maggiore} quantum field theory makes no prediction
about the actual value of $\langle 0\vert T_{\mu \nu}\vert 0\rangle_{\text{ren}}$, just in the same way as it does
not predict the physical (renormalized) value of the electron charge or mass, but these
quantities can only be obtained from experimental measurements.
Despite the fact that the physical value of the vacuum energy can
only be determined from observations, there are, however, several conditions that
 from a purely phenomenological point of view (and neglecting possible fine-tuned cancellations
with other contributions) a physical renormalized
energy-momentum tensor should satisfy
\begin{itemize}
\item $\langle 0\vert T_{\mu \nu}\vert 0\rangle_{\text{ren}}$ should be covariant and conserved.
Notice that in a curved space-time a covariant expectation value of the energy-momentum tensor
can involve not only  the metric tensor but also other tensors such as curvatures  or any other object
which transforms covariantly under diffeomorphisms \cite{Birrell}. 
On the other hand, regarding conservation, in general, if nongravitational interactions of the scalar field  are taken into
account, then vacuum energy could be coupled to other components and the  conservation
should be required for the total energy-momentum tensor. In any case, in this work we
limit ourselves to the simplest noninteracting case and therefore the vacuum energy-momentum
tensor should be independently conserved. 

\item $\rho_{\text{ren}}\lsim \rho_c$, i.e. in
order to have phenomenologically viable contributions,  the
vacuum energy should be smaller  than the dominant component of the
Universe at early times. Only at late times, and if we assume vacuum energy  to play
a role in the dark matter or dark energy problems, its value could be
comparable to the critical density $\rho_c=3H^2/8\pi G$.
\end{itemize}

In this work we will explore the possibility of constructing the
renormalized vacuum energy momentum by means of a {\it comoving} three-momentum cutoff.
Unlike previous works which focused on {\it physical} cutoff scales, the use of this kind
of regularization provides covariant expressions for the regularized integrals and also
guarantees that the bare energy-momentum tensor is conserved. Accordingly, we do not need to include
noncovariant counterterms in order to render the final results covariant. This can easily be seen  in the following example.  Let us consider the cutoff regularized bare energy-momentum
tensor for minimally coupled massless scalar fields in a Robertson-Walker background \cite{Parker, Fulling, Maggiore}. In this case
\begin{eqnarray}
\rho_{\text{bare}}&=&\frac{\Lambda_P^4}{16\pi^2}+\frac{H^2(t)\Lambda_P^2}{16\pi^2}+\Od(H^4\ln \Lambda_P) \\
p_{\text{bare}}&=&\frac{\Lambda_P^4}{48\pi^2}+c_1\frac{H^2(t)\Lambda_P^2}{16\pi^2}+\Od(H^4\ln \Lambda_P)
\end{eqnarray}
where $c_1=$ $-1/3$, 1, 2/3 in the de Sitter, radiation and matter eras respectively. Here $\Lambda_P$
is a constant physical momentum cutoff. Notice that indeed the use  of the physical cutoff prevents the bare energy-momentum tensor from
being conserved. This is clearly seen for example from the dominant  quartic terms whose effective
equation of state would be $p_{\text{bare}}=\frac{1}{3}\rho_{\text{bare}}$, i.e. corresponding
to radiation, but, however, they do not scale with $a(t)$. As shown in \cite{Maggiore}
 this is not a problem, since the bare quantities are not  observable and
by adding appropriate (noncovariant) counterterms it would always be possible to render the
renormalized energy-momentum tensor  conserved, provided $\rho_{\text{ren}}=-p_{\text{ren}}$.

However, if we consider instead a constant comoving cutoff $\Lambda_c$, the above results read
\begin{eqnarray}
\rho_{\text{bare}}&=&\frac{\Lambda_c^4}{16\pi^2a^4}+\frac{H^2(t)\Lambda_c^2}{16\pi^2a^2}+\Od(H^4\ln\Lambda_c)
\label{bare1} \\
p_{\text{bare}}&=&\frac{\Lambda_c^4}{48\pi^2a^4}+c_1\frac{H^2(t)\Lambda_c^2}{16\pi^2a^2}+\Od(H^4\ln \Lambda_c)
\label{bare2}
\end{eqnarray}
which yield  a conserved bare energy-momentum tensor as expected
according to our previous discussion. Indeed, notice that now the leading
quartic term scales as expected according to its
equation of state, i.e. as radiation, and the same is true for
the rest of  terms.

Since each of the divergent contributions (quartic, quadratic or logarithmic)
is conserved independently, it would be possible in principle
to add different conserved counterterms
for each of them. In the simplest possibility,  the counterterms are just given by the
same expressions (\ref{bare1}) and (\ref{bare2})
but in which the modes have been integrated
from some constant comoving renormalization scale $\Lambda_R$ up to the ultraviolet cutoff $\Lambda_c$, i.e.
just  subtracting the contributions of the modes in the range $[\Lambda_R,\Lambda_c]$.
Then in this case the physical interpretation of the renormalized quantities is
straightforward, since only the modes in the unsubtracted range $[0,\Lambda_R]$
will contribute. Thus, we can simply write
     \begin{eqnarray}
\rho_{\text{ren}}&=&\frac{\Lambda_R^4}{16\pi^2a^4}+\frac{H^2(t)\Lambda_R^2}{16\pi^2a^2}+\Od(H^4\ln \Lambda_R) \\
p_{\text{ren}}&=&\frac{\Lambda_R^4}{48\pi^2a^4}+c_1\frac{H^2(t)\Lambda_R^2}{16\pi^2a^2}+\Od(H^4\ln \Lambda_R)\,.
\end{eqnarray}

Notice that $\Lambda_R$  is understood as a limit on the frequency of the Fourier modes
which actually contribute to the vacuum energy and in general can 
be different from the standard quantum field theory UV cutoff which sets the range of validity of the 
theory. Thus for instance in \cite{holo2,holo3}
$\Lambda_R$ is obtained by demanding that no state in the Hilbert space can have an 
energy such that the corresponding Schwarzschild radius exceeds the Hubble (or the event) horizon, 
i.e. modes which would have collapsed in a black hole are excluded in the computation 
of the vacuum energy. 
This is a generic prediction of so called holographic \cite{holo1}
but also of nonholographic \cite{Visser} entropy bounds on the number of physical quantum
states for any gravitating system. These models generically 
predict $\Lambda_R$ much smaller than the quantum field theory cutoff, thus
alleviating the cosmological constant problem. 
In this work, however, we will not assume any particular scenario for the determination of $\Lambda_R$,
instead we will adopt a phenomenological point of view leaving it  as a free parameter 
to be fixed by observations. 

When  trying to extend the renormalization procedure we have just 
described  to  the case of
massive fields,  an additional scale $m$ appears in the problem which
opens up different regimes for the vacuum energy behavior. Thus,
in the case in which the comoving mass is larger than the renormalization scale, i.e.
$m^2a^2> \Lambda_R^2$, we will show  that the effective equation of state of vacuum energy is
that of nonrelativistic matter.
In order to obtain this kind of results,  it will be necessary to  determine the behavior
of the integrals not only in the high-momenta (UV) regime, as is usually considered in the literature, but
also for low momenta (IR). We will show that the dark matter behavior of vacuum energy
also holds at the level of perturbations, thus opening up the quite unexpected possibility
for the vacuum energy to  form  large scale structures. 

The paper is organized as follows: in Sec.\ II we introduce the basic expressions for the quantization of
scalar fields in a Robertson-Walker background. In Sec.\ III we particularize to de Sitter
space-times and obtain exact expressions for both massless and massive fields. In Sec.\ IV
we consider asymptotic expressions in the UV limit, and in Sec.\ V the results in the IR are
discussed in more detail. Section VI is devoted to the generalization to arbitrary Robertson-Walker
geometries, and in Sec.\ VII we calculate the vacuum energy-momentum tensor on
a perturbed Robertson-Walker background and obtain the general expression for the evolution
of the density contrast of the vacuum energy. Section VIII contains the main conclusions of the work.


\section{Scalar fields in Robertson-Walker backgrounds}


Let us consider a scalar field $\phi$ with mass $m$ in a spatially flat Robertson-Walker background 
\mbox{$\text{d}s^2=a^2(\eta)(\text{d}\eta^2-\text{d}{\bf x}^2)$}. The corresponding Klein-Gordon equation reads
\begin{eqnarray}
\Box \, \phi +m^2\, \phi = 0 \,.
\label{KG}
\end{eqnarray}
Thus, the field $\phi({\bf x},\eta)$ can be Fourier expanded as
\begin{eqnarray}
\phi ({\bf x},\eta)=\int \text{d}^3{\bf k}\, \left(a_{{\bf k}}\, \phi_{k}(\eta)\,e^{i\, {\bf k}\,{\bf x}} + a^{\dag}_{{\bf k}}\, \phi^*_{k}(\eta)\,e^{-i\, {\bf k}\,{\bf x}}\right)\,.
\label{fielddecomposition}
\end{eqnarray}
The scalar field can be quantized by letting $a_{{\bf k}}$ and $a^{\dag}_{{\bf k}}$ be operators which satisfy the usual commutation relations
\begin{eqnarray}
[a_{{\bf p}},a^{\dag}_{{\bf q}}]=\delta^{(3)}({\bf p}-{\bf q})\,.
\end{eqnarray}
Introducing $\psi_{k}$ by
\begin{eqnarray}
\phi_{k} = \frac{\psi_{k}}{a}\,,
\label{mode}
\end{eqnarray}
Eq.\ \eqref{KG} can be recast as
\begin{eqnarray}
\psi_{k}'' + \left(k^2 - \frac{a''}{a} + m^2 a^2 \right)\psi_{k}=0\,.
\label{fieldequation}
\end{eqnarray}

It can be shown \cite{Parker,Fulling} that the mean value of the off-diagonal elements of the energy-momentum tensor for the vacuum state $|0\rangle$ of a scalar field are zero as expected from symmetry considerations while the diagonal elements, i.e.\ the energy density and pressure, are
\begin{eqnarray}
\rho=\langle 0\vert T^{0}_{\;0}\vert 0\rangle&=& \int \frac{\text{d}^3{\bf k}}{2a^2} \left(\vert{\phi'_{k}}\vert^2+k^2\vert\phi_{k}\vert^2
+m^2a^2\vert\phi_{k}|^2\right)\nonumber
\\
\end{eqnarray}
\begin{eqnarray}
p=-\langle 0\vert T^{i}_{\;i}\vert 0\rangle&=& \int \frac{\text{d}^3{\bf k}}{2a^2} \left(\vert{\phi'_{k}}\vert^2-\frac{ k^2}{3}\vert\phi_{k}\vert^2 
-  m^2a^2\vert\phi_{k}|^2\right)\,.\nonumber \\
\end{eqnarray}
Therefore according to the discussion in the Introduction, for the renormalized quantities 
we will consider
\begin{eqnarray}
\rho_{\text{ren}}&=& \frac{2\pi}{a^2}\int_0^{\Lambda_R} \text{d}{k \,k^2} \left(\vert{\phi'_{k}}\vert^2+k^2\vert\phi_{k}\vert^2
+m^2a^2\vert\phi_{k}|^2\right)\nonumber
\\
\label{energydensity}
\end{eqnarray}
\begin{eqnarray}
p_{\text{ren}}&=&  \frac{2\pi}{a^2}\int_0^{\Lambda_R} \text{d}{k \,k^2} \left(\vert{\phi'_{k}}\vert^2-\frac{ k^2}{3}\vert\phi_{k}\vert^2 
-  m^2a^2\vert\phi_{k}|^2\right)\,.\nonumber \\
\label{pressure}
\end{eqnarray}

 We shall be interested in the equation of state satisfied by the energy density and pressure of the vacuum state of a scalar field $w=p_{\text{ren}}/\rho_{\text{ren}}$. From Eqs.\ \eqref{energydensity} and \eqref{pressure} it is seen that if the kinetic term of the field, the gradient term or the mass term dominates one gets for $w$ the values $1$, $-1/3$ or $-1$ respectively. Furthermore, if the mass term is negligible with the 
 kinetic and gradient terms  of the same order of magnitude (as it happens for instance
in UV regime $k\rightarrow \infty$) then $w=1/3$, i.e we get a radiation behavior. Finally, if the kinetic and mass terms are of the same order of magnitude, with the gradient term negligible, one expects $w=0$. Hence, for a massive field we expect a matter contribution to the vacuum energy density of the field in the IR region. We shall perform numerical calculations to test these qualitative considerations using exact analytical expressions derived in the next section.

\section{Exact solutions in de Sitter space-time}

In this particular case the scale factor reads
\begin{eqnarray}
a(\eta)=-\frac{1}{H\eta}\,
\label{aconformaldesitter}
\end{eqnarray}
with $H$ the constant Hubble parameter. The range of the conformal time is $-\infty<\eta<0$. 
The Klein-Gordon equation \eqref{fieldequation} results in
\begin{eqnarray}
\psi_{k}'' + \left[k^2 -\left(2-\frac{m^2}{H^2}\right) \frac{1}{\eta^2} \right]\psi_{k}=0\,.
\label{fieldequationdesitter}
\end{eqnarray}

%


\subsection{Massless scalar field}

\begin{figure}

    \begin{center}
        {\includegraphics[width=0.47\textwidth]{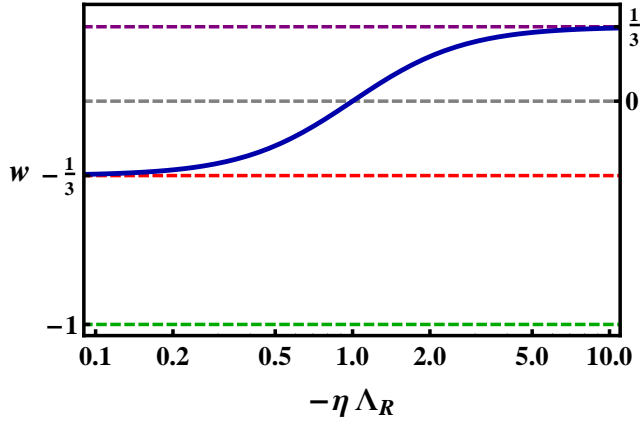}}
    \caption {\footnotesize Evolution of the equation of state parameter $w$ for
massless fields in terms of the conformal time $\eta$ and the renormalization cutoff $\Lambda_{R}$ (in linear-log scale). For past times or high cutoff the dominant contribution is the radiation one, $w=1/3$, and for the symptotic future or low cutoff $w=-1/3$.
    }
    \label{fig:dustmasaless}
    \end{center}
\end{figure}

As a particular case, we briefly illustrate the calculation for a massless field. The field equation \eqref{fieldequationdesitter} for $m=0$ is
\begin{eqnarray}
\psi_{k}'' + \left(k^2 -\frac{2}{\eta^2} \right)\psi_{k}=0\,,
\label{fielequationdSm0}
\end{eqnarray}
whose positive frequency solution is given by
\begin{eqnarray}
\psi_{k}(\eta)=\frac{1}{(2\pi)^{3/2}} \frac{1}{\sqrt{2k}}\left(1-\frac{i}{k\eta} \right) e^{-ik\eta}\,
\label{solutionm0}
\end{eqnarray}
where the normalization has been imposed in order to match plane waves in the UV ($\psi_{k} \sim e^{-i k \eta}/\sqrt{(2\pi)^3 2k}$ when $-k \eta \rightarrow \infty$).
Let us recall that by taking the positive frequency solution, a choice of the vacuum state is made.
The mode functions of the field \eqref{mode} are
\begin{eqnarray}
\phi_{k}(\eta)=
-\frac{1}{(2\pi)^{3/2}} \frac{H}{\sqrt{2k}}\left(\eta-\frac{i}{k} \right) e^{-ik\eta}\,.
\end{eqnarray}
Therefore, the renormalized energy density and pressure are
\begin{widetext}
\begin{eqnarray}
\rho^{m=0}_{\text{dS,ren}} = \frac{H^2\eta^2}{2} \int \text{d}^3{\bf k} \left(|{\phi'_{k}}|^2+  k^2  |\phi_{k}|^2 \right) =  \frac{H^4}{4\pi^2}\int_0^{\Lambda_R} \text{d}k\left( k^3 \eta^4+ \frac{k\eta^2}{2} \right)
\label{energydensitym0dS}
\end{eqnarray}
\begin{eqnarray}
p^{m=0}_{\text{dS,ren}}=\frac{H^2\eta^2}{2}\int \text{d}^3{\bf k} \left(|{\phi'_{k}}|^2-\frac{ k^2}{3} |\phi_{k}|^2 \right) = \frac{H^4}{4\pi^2}\int_0^{\Lambda_R} \text{d}k \left( \frac{1}{3} k^3 \eta^4 -\frac{k\eta^2}{6} \right) \,.
\label{pressurem0dS}
\end{eqnarray}
\end{widetext}
It is seen that the energy density has two contributions. Since the dependence on time will remain the same after performing the integration (thanks to the constancy of the comoving cutoff $\Lambda_R$), it is possible to state that one of them evolves in time like radiation $\propto \eta^{4}$ with 
effective equation of state $w=1/3$
 and the other one  $\propto \eta^2$ (or equivalently $\propto 1/a^2$), as
in a curvature dominated universe with $w=-1/3$. Notice that the 
scaling of the different terms of the energy density agree with the corresponding ratios $p/\rho$ which
implies that the renormalized energy-momentum tensor is conserved. Let us recall that the dependence of $w$ on time and on the renormalization cutoff is through the product $-\eta \Lambda_{R}$. 
In Fig.\ \ref{fig:dustmasaless} the dependence of the equation of state coefficient $w$ on time and the renormalization cutoff is plotted.

\subsection{Massive scalar field}


For the case where $m \not= 0 $, we must deal with the general equation \eqref{fieldequationdesitter}, which may be put in the following form \cite{Riotto}
\begin{eqnarray}
\psi_{k}'' + \left(k^2 - \frac{\nu^2-\frac{1}{4}}{\eta^2} \right)\psi_{k}=0\,
\label{Bessel}
\end{eqnarray}
where
\begin{eqnarray}
\nu^2=\frac{9}{4}-\frac{m^2}{H^2}\,.
\label{numasa}
\end{eqnarray}

\begin{figure}
    \begin{center}
        {\includegraphics[width=0.47\textwidth]{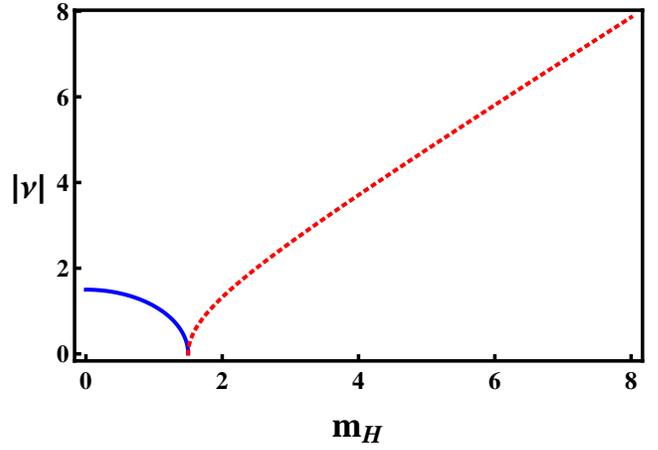}}
    \caption {\footnotesize Dependence of the modulus of $\nu$ on $m_{H}$. In the range where the line is blue and continuous $\nu$ is real, then $\nu$ turns to be pure imaginary in the red dashed line range. For $m_{H} = 0 \Rightarrow \nu=3/2$, if $m_{H}$ increases $\nu$ decreases until zero when $m_{H}=3/2$. Then $\nu$ turns out to be purely imaginary increasing in modulus when $m_{H}$ increases. For $m_{H}\gg 1$, $\nu \approx i m_{H}$ (for instance, for $m_{H}=5$ their absolute difference is less than $5\%$).
    }
    \label{fig:masanu}
    \end{center}
\end{figure}

Let us recall that from now on the dependence on the mass of the scalar field will be encoded in $\nu$. Let us define a dimensionless parameter $m_{H}$ as
\begin{eqnarray}
m_{H}\equiv \frac{m}{H}\,.
\label{mH}
\end{eqnarray}
 We depict in Fig.\ \ref{fig:masanu} the dependence of the modulus of $\nu$ on $m_{H}$.

Equation \eqref{Bessel} is of the Bessel type and has as a general solution
\begin{eqnarray}
\psi_{k}(\eta)=\sqrt{-\eta}\left[c_{1}(k) H^{(1)}_{\nu}(-k\eta) + c_{2}(k) H^{(2)}_{\nu}(-k\eta) \right]\,, \nonumber \\
\end{eqnarray}
where $H^{(i)}_{\nu}(x)$ are Bessel functions of the third kind also called Hankel functions. Let us remark that this solution is still valid when $\nu \in \mathbb{C}$, i.e. $m_{H}>3/2$.
In order to impose that this solution matches in the UV domain with a plane wave ($e^{-ik\eta}/\sqrt{2k}$), which are the expected solutions in flat space-time, the coefficients $c_{1}(k)$ and $c_{2}(k)$ must be
\begin{eqnarray}
c_{1}(k)=\frac{1}{(2\pi)^{3/2}} \frac{\sqrt{\pi}}{2}\, e^{i\left(\nu+\frac{1}{2}\right)\frac{\pi}{2}}\ \ \ \ \ \ \ \ c_{2}(k)=0\,,
\label{cs}
\end{eqnarray}
where the behavior of the Hankel functions for real values of the argument tending to infinity has been taken into account.\footnote{ The asymptotic behavior of the Hankel functions is
\begin{eqnarray}
H^{(1)}_{\nu}(x) &\sim & \sqrt{\frac{2}{\pi x}}\, e^{i\left(x-\frac{\nu \pi}{2}-\frac{\pi}{4}\right)}\,\nonumber\\
 H^{(2)}_{\nu}(x) &\sim & \sqrt{\frac{2}{\pi x}} \, e^{-i\left(x-\frac{\nu \pi}{2}-\frac{\pi}{4}\right)}\,. \nonumber
\end{eqnarray}}
The choice of the particular coefficients \eqref{cs} is equivalent to a choice of the vacuum state \cite{Birrell,Maggiore}.
Thus, the solution of \eqref{Bessel} we will focus on may be recast into
\begin{eqnarray}
\psi_{k}(\eta)=- \frac{1}{(2\pi)^{3/2}} \Theta_{\nu} \sqrt{k}\, \eta \, h^{(1)}_{\nu-1/2}(-k\eta)\,
\label{solutionmspherical}
\end{eqnarray}
where $h^{(1)}_{\nu-1/2}(x)$ is the spherical Hankel function of the first order
\begin{eqnarray}
h^{(1)}_{n}(x)=\sqrt{\frac{\pi}{2x}}\, H^{(1)}_{n+1/2}(x)\,,
\end{eqnarray}
and in order to simplify the notation we have defined
\begin{eqnarray}
\Theta_{\nu}=\frac{1}{\sqrt{2}}\, e^{i\left(\nu+\frac{1}{2}\right)\frac{\pi}{2}}\,.
\end{eqnarray}
To express the solution in terms of spherical Hankel functions simplifies the calculations. The spherical Hankel functions of integer order are polynomials in $1/x$ multiplied by a phase $e^{i x}$, which in fact gives straightforward expressions when $\nu$ takes half-integer values. For instance, when $\nu=3/2$ ($m=0$) we obtain
\begin{eqnarray}
\psi^{m=0}_{k}(\eta)=\frac{1}{(2\pi)^{3/2}} \sqrt{\frac{k}{2}}\, \eta \, h^{(1)}_{1}(-k\eta)\,.
\end{eqnarray}
Since
\begin{eqnarray}
h^{(1)}_{1}(x)=-\left(\frac{1}{x}+\frac{i}{x^2}\right) e^{ i x}\,,
\end{eqnarray}
we get
\begin{eqnarray}
\psi^{m=0}_{k}(\eta) =\frac{1}{(2\pi)^{3/2}} \frac{1}{\sqrt{2k}}\left(1-\frac{i}{k\eta} \right) e^{-ik\eta}\,
\end{eqnarray}
which is in accordance with \eqref{solutionm0}.

Therefore, the mode functions $\phi_{k}$ \eqref{mode} of the field are
\begin{eqnarray}
\phi_{k}(\eta)=
\frac{1}{(2\pi)^{3/2}} \Theta_{\nu}\, H\, \sqrt{k}\,  \eta^2 \, h^{(1)}_{\nu-1/2}(-k\eta)\,,
\label{solution}
\end{eqnarray}
where we have used the expression for the scale factor in a de Sitter stage \eqref{aconformaldesitter}.
Since there is no risk of confusion, from now on the Hankel functions $H^{(1)}_{\nu}$ and $h^{(1)}_{\nu-1/2}$ will be denoted as $H_{\nu}$ and $h_{\nu-1/2}$ respectively.

\subsubsection{Energy density and pressure}


The energy density $\rho_{\text{dS}}$ and pressure $p_{\text{dS}}$ of a massive scalar field in a de Sitter stage is calculated from the expressions
\begin{eqnarray}
\label{deSitter}
\rho_{\text{dS}} =
\frac{H^2\eta^2}{2} \int \text{d}^3{\bf k} \left(|{\phi'_{k}}|^2+  k^2 |\phi_{k}|^2 + m^2 |\phi_{k}|^2\right) \nonumber \\ \ \\
p_{\text{dS}} = \frac{H^2\eta^2}{2} \int \text{d}^3{\bf k} \left(|{\phi'_{k}}|^2- \frac{k^2 }{3} |\phi_{k}|^2 - m^2 |\phi_{k}|^2\right) \nonumber \,.
\end{eqnarray}
The exact analytical expressions are presented in the Appendix.
Let us recall that depending on which term or pair of terms dominates different values for $w$ are
expected, in particular $1,-1/3,-1,1/3,0$ and $-2/3$.

\subsection{Massive scalar field with $m_{H} = \sqrt{2}$}


There is a particular simple case when $\nu = 1/2 $, i.e. when $m_{H} = \sqrt{2}$. In this case, the equation for $\psi_k$ reduces to that associated with a massless
field in a flat space-time
\begin{eqnarray}
\psi_{k}'' + k^2\,\psi_{k}=0\,.
\label{Flat}
\end{eqnarray}
Thus, the general Bessel equation \eqref{Bessel} has the simple plane wave solution
\begin{eqnarray}
\psi_{k}(\eta)=\frac{1}{(2\pi)^{3/2}}\frac{1}{\sqrt{2k}}e^{-ik\eta}\,,
\end{eqnarray}
where we have already imposed the standard normalization in flat space-time.
The energy density $\rho^{\nu = 1/2}_{\text{dS}}$ and pressure $p^{\nu = 1/2}_{\text{dS}}$ 
can be computed by using Eq. (\ref{deSitter}) and by taking into account the relation 
(\ref{mode}) between $\phi_k$ and $\psi_{k}$. In this case, the result
can be written in a simple form
\begin{eqnarray}
\rho^{\nu = 1/2}_{\text{dS,ren}} = \frac{H^4}{8 \pi^2 } \int^{\Lambda_R}_{0} \text{d} k  \left( 2\, k^3\, \eta^4   + 3 k\, \eta^2 \right)\,
\label{energydensity1o2}
\end{eqnarray}
\begin{eqnarray}
p^{\nu = 1/2}_{\text{dS,ren}} = \frac{H^4}{8 \pi^2 } \int^{\Lambda_R}_{0} \text{d} k \left(\frac{2}{3} \,k^3\, \eta^4 -k\, \eta^2 \right)\,.
\label{pressure1o2}
\end{eqnarray}
As it happens for the massless case, the energy density has two simple contributions:
one behaves as radiation and it is proportional to $\eta^4$, whereas the other one is
proportional to  $\eta^2$.

The reason for this massless behavior for this particular mass is easy to understand. In a de Sitter space-time a minimally coupled scalar field with mass $m_H$ obeys the same equation that a conformal coupled field with mass $m^2_{H,\text{conformal}}=m_H^2-2$. The two different couplings are accounted for by a redefinition of the field mass. Therefore, $m_H=\sqrt{2}$ corresponds to the massless conformal field.


\section{UV regime}


It is interesting to study the behavior of the energy density  \eqref{energydensitydS} in the limit 
$\Lambda_R\gg ma$. In this case, the energy density is dominated by the large $k$ Fourier modes (UV domain). This may easily be done by using the asymptotic expansion for the spherical Hankel functions.
We state the results
\begin{widetext}
\begin{eqnarray}
\rho^{UV}_{\text{dS,ren}} \simeq \frac{H^4}{8 \pi^2 } \int^{\Lambda_R}_0 \text{d} k  \left( 2\, k^3\, \eta^4   + \left(1+m_{H}^2 \right) k\, \eta^2 -\frac{m_{H}^2 \left(m_{H}^2-2\right)}{4\, k } \right)\,
\label{energydensitydSasymptotic}
\end{eqnarray}
\begin{eqnarray}
p^{UV}_{\text{dS,ren}} \simeq \frac{H^4}{8 \pi^2 } \int^{\Lambda_R}_{0} \text{d} k \left(\frac{2}{3} \,k^3\, \eta^4 - \frac{\left(1+m_{H}^2\right)}{3}\, k\, \eta^2 + \frac{m_{H}^2 \left(m_{H}^2-2\right)}{4\, k} \right)\,,
\label{pressuredSasymptotic}
\end{eqnarray}
\end{widetext}
where in these  expressions only the asymptotically large  contributions (power-law and logarithmic in $\Lambda_R$) in the UV are retained. Notice that these are the kind of terms usually considered in the literature when studying contributions to the zero-point energy. Therefore, in the UV domain the energy density has three contributions that evolve in time like radiation $\propto \eta^4$, $\propto \eta^2$ and a cosmological constant. Furthermore, it is readily seen from the pressure expression that these terms obey equations of state with $w_{\eta^4}=1/3$, $w_{\eta^2}=-1/3$ and $w_{\eta^0}=-1$ respectively. It is to be noted that again the evolution in time fits with the corresponding equation of state, which 
implies that the renormalized energy-momentum tensor is conserved. Hence, the vacuum energy density of a massive scalar field in the UV behaves as a radiation fluid, a $w=-1/3$ fluid and a cosmological constant. Notice that any of these contributions could be removed by choosing appropriate
counterterms as discussed in the Introduction.

The radiation term does not depend on the mass while the other ones increase in absolute value as the mass increases. It is seen that for a massless field $m_{H}=0$ or a massless conformal field $m_{H}=\sqrt{2}$ the cosmological constant contribution vanishes. This result for a massless field is also obtained using a dimensional regularization scheme \cite{Akhmedov} in a flat space-time. However, the second case is a proper feature of the curved space-time we are considering, where the minimal and conformal couplings differ by a redefinition of the field mass. The term that evolves as $\eta^2$ does not vanish for any physical value of $m_{H}$.


\section{IR regime}

As time evolves, the UV condition  $\Lambda_R\gg ma$ would be violated and 
the approximation used in the previous section would no longer be valid. Thus, at 
sufficiently late times (or for large enough masses) we are in the IR regime $\Lambda_R\ll ma$. 
For the leading contribution with $k^2\ll -\nu^2/\eta^2$, we obtain from the equation of motion  (\ref{Bessel})
\begin{eqnarray}
\psi_{k}'' + \frac{m_H^2}{\eta^2}\psi_{k}=0\,,
\end{eqnarray}
whose solution can be written as
\begin{eqnarray}
\psi_{k}(\eta)=\frac{c_1(k)}{\sqrt{2ma}} \,e^{-i m_H\int^\eta\frac{\text{d}\eta'}{\eta'}}+\frac{c_2(k)}{\sqrt{2ma}}\, e^{i m_H\int^\eta\frac{\text{d}\eta'}{\eta'}}\,.
\end{eqnarray}
Thus, the positive frequency solution with the correct normalization reads
\begin{eqnarray}
\psi_{k}(\eta)=
\frac{1}{(2\pi)^{3/2}\sqrt{2ma}}\,
e^{-i m_H\int^\eta\frac{\text{d}\eta'}{\eta'}}
\end{eqnarray}
and the corresponding renormalized energy density and pressure read to leading order
\begin{eqnarray}
\rho^{IR}_{\text{dS,ren}}&\simeq& \int\frac{\text{d}^3{\bf k}}{2(2\pi)^3a^4}ma=
\frac{H^4}{4\pi^2} \int_0^{\Lambda_R} \text{d}k\; m_Hk^2\eta^3\nonumber \\
p^{IR}_{\text{dS,ren}}&\simeq & 0\,.
\end{eqnarray}
Thus, we see that the leading contribution of the vacuum energy in the IR regime corresponds
to nonrelativistic matter which scales as $\rho\propto a^{-3}$ and the corresponding 
equation of state is indeed $w=0$ to this order. Notice that because of the growth of 
$a(\eta)$, we will have  $\Lambda_R\ll ma$ at some future time
 for any cutoff $\Lambda_R$; i.e. the matter behavior
will be a  generic prediction at late times. Notice also, that this matter contribution
 was not present in the UV regime, so that we expect it to be a genuine IR effect 
not affected by the way in which the UV modes are regularized. 

As time evolves, we expect a transition in the equation of state of vacuum energy
from radiation when UV modes dominate at early times, to a matter domination when the IR 
modes dominate at late times. In order to explicitly explore this behavior we have 
evaluated numerically the equation of state parameter $w$  as a function of time.

\begin{figure}
    \begin{center}
        {\includegraphics[width=0.47\textwidth]{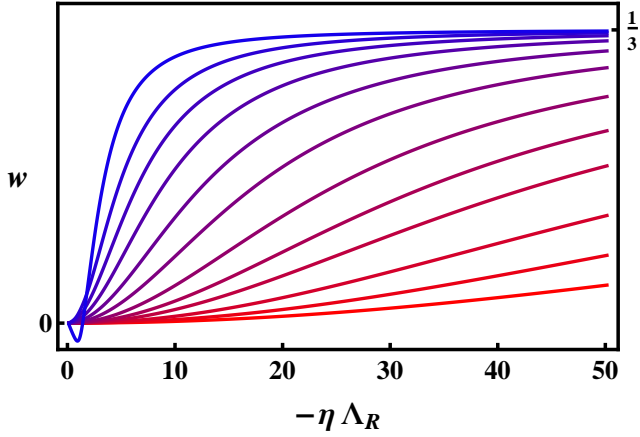}}
    \caption {\footnotesize Evolution of the equation of state parameter $w$ in terms of the conformal time $\eta$ and the renormalization cutoff $\Lambda_{R}$ for several values of the field mass $m_{H}$ (from top to bottom $m_{H}=2, 3.5, 5, 7, 10, 14, 20, 27, 35, 50, 70, 100$) (in linear-linear scale). For past times (or high cutoff) the dominant contribution is the radiation one $w=1/3$. As the time goes by (or for low cutoff), a matter behavior $w=0$ appears.  Moreover, when the mass of the field increases the behavior as matter spreads in time (or in momentum space).}
    \label{fig:dustmasa}
    \end{center}
\end{figure}

In Fig.\ \ref{fig:dustmasa},  $w$  for a massive scalar field in a de Sitter universe is depicted as a function of the conformal time $\eta$ and the renormalization cutoff $\Lambda_{R}$ for several values of the mass $m_{H}$ of the field ($0<m_{H}<100$). It is found  that when $\eta$ tends to zero (or for a low cutoff), a matter behavior is found 
which spreads in time (or in momentum space) for bigger masses in accordance with the physical interpretations (bigger mass implies a bigger matter contribution to the energy-momentum tensor). This matter behavior is present in the IR region since it is not observed in the asymptotic expansion \eqref{energydensitydSasymptotic} and \eqref{pressuredSasymptotic}.

\begin{figure}
    \begin{center}
        {\includegraphics[width=0.47\textwidth]{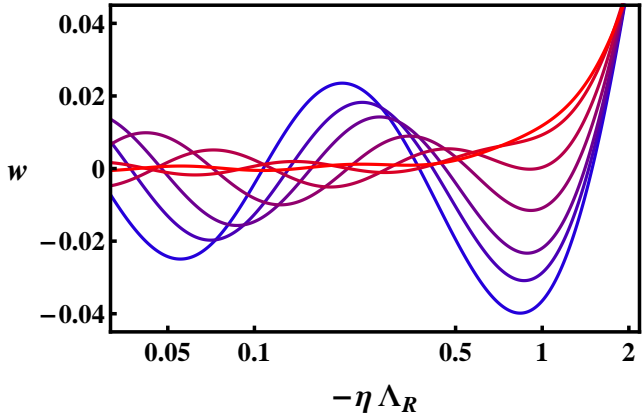}}
    \caption {\footnotesize Oscillations observed in the evolution of the equation of state parameter $w$ for field masses $m_{H}$ in the range $1.9<m_{H}<2.75$ (from blue to red lines $m_{H}=1.9, 1.95, 2, 2.1, 2.25, 2.5, 2.75$). The plot is in linear-log scale.
    }
    \label{fig:dustmasaosc}
    \end{center}
\end{figure}

Oscillations of  $w$  are observed for a field mass $m_{H}=2$ in Fig.\ \ref{fig:dustmasa}. A more detailed analysis of the range of masses $1.9<m_{H}<2.75$ is shown in Fig.\ \ref{fig:dustmasaosc}. These oscillations are present in this range of masses and they damp as the mass increases, being
negligible for $m_{H}>3$.

\begin{figure}
    \begin{center}
        {\includegraphics[width=0.47\textwidth]{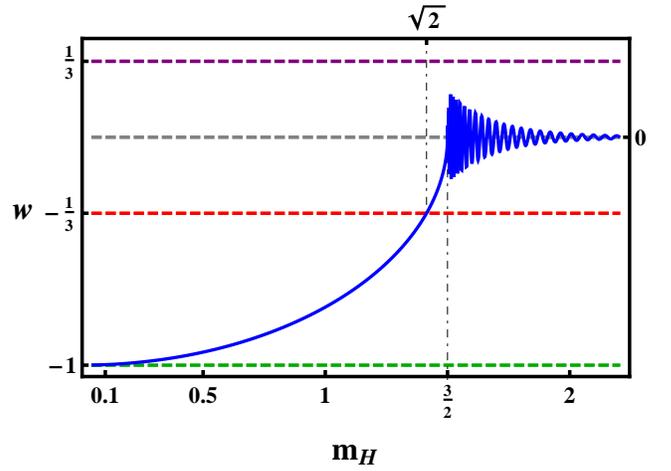}}
    \caption {\footnotesize Linear-scale plot of the equation of state parameter $w$ when $-\eta\,\Lambda_{R}=0.1$ for field masses in the range $0<m_{H}<2.2$. For small masses the dominant contribution is a cosmological constant, and then as the mass increases $w$ tends to zero. When $m_{H}=\sqrt{2}$, the cosmological constant contribution vanishes as it is seen in the exact solution \eqref{energydensity1o2} and \eqref{pressure1o2}, as well as in the UV approximation \eqref{energydensitydSasymptotic} and \eqref{pressuredSasymptotic}.
}
    \label{fig:futuremasa}
    \end{center}
\end{figure}

\begin{figure}
    \begin{center}
        {\includegraphics[width=0.47\textwidth]{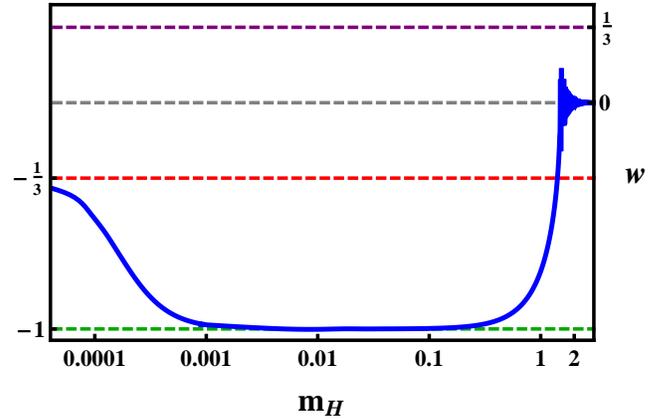}}
    \caption {\footnotesize Linear-log scale plot of the equation of state parameter $w$ when $-\eta\,\Lambda_{R}=0.1$ for field masses in the range $3\times 10^{-5}<m_{H}<2.2$. The dominant contribution of the massless case ($w=-1/3$) is smoothly recovered as we reduce the mass of the field.
}
    \label{fig:futuremasapeq}
    \end{center}
\end{figure}

For a fixed time $\eta$ and renormalization cutoff $\Lambda_{R}$, there is a transition for increasing 
values of the mass from the massless case, when the $w=-1/3$ term dominates, passing through a cosmological constant phase for low masses until the dust dominated case $w=0$ for high masses. We depicted in Fig.\ \ref{fig:futuremasa} the equation of state parameter $w$ when $-\eta\,\Lambda_{R}=0.1$ for several values of the field mass in the range $0<m_{H}<2.2$.
 Let us recall that when $m_{H}=\sqrt{2}$ the cosmological constant term disappears in the UV domain as deduced from the exact solution for this case \eqref{energydensity1o2} and \eqref{pressure1o2}, and from the asymptotic expansion for the energy density \eqref{energydensitydSasymptotic} and pressure \eqref{pressuredSasymptotic}. When $m_{H}=0$ we must recover $w=-1/3$ in the future. From the previous figure it does not seem so. However, if we perform a finer calculation near zero masses we obtain the results depicted in Fig.\ \ref{fig:futuremasapeq} where the $w=-1/3$ behavior of a massless field smoothly changes  to $w=-1$.
%


\section{Arbitrary Robertson-Walker geometries}


In the previous sections, we found that the vacuum energy of sufficiently heavy fields in a de Sitter background behaves as nonrelativistic matter at late times.  In the following we generalize this result for arbitrary Robertson-Walker backgrounds.

Considering the case $ma\gg \Lambda_R$ and assuming $m\gg H$ at late times (as is
indeed the case for  
Standard Model particle masses), Eq.\ \eqref{fieldequation} is reduced to
\begin{eqnarray}
\psi_{k}'' + m^2 a^2 \psi_{k}=0\,.
\end{eqnarray}
Hence, in this
limit the following approximated solution can be obtained by the WKB method
\begin{eqnarray}
\phi _k(\eta)&=& \frac{c_{1}(k)}{\sqrt{2ma^3}} \exp{\left(-im\int a \,\text{d}\eta\right)}\nonumber \\
&+&\frac{c_{2}(k)}{\sqrt{2ma^3}} \exp{\left(im\int a\, \text{d}\eta\right)}\,.
\end{eqnarray}
Notice that in this limit all the $k$ modes evolve in time in the same way. Therefore, if we consider the positive frequency solution and normalize it according to \eqref{scalar} [$c_1=1/(2\pi)^{3/2}$, $c_2=0$], we can  calculate the energy density and pressure from Eqs.\ \eqref{energydensity} and \eqref{pressure}
\begin{eqnarray}
\rho_{\text{ren}} = \frac{1}{8\pi^2} \int_0^{\Lambda_R} \text{d}k \,k^2\left( \frac{2m}{a^3}+\frac{9H^2}{4ma^3} + \frac{k^2}{ma^5} \right)
\label{energydensityarbitrary}
\end{eqnarray}
\begin{eqnarray}
p_{\text{ren}} =  \frac{1}{8\pi^2} \int_0^{\Lambda_R} \text{d}k \,k^2\left( \frac{9 H^2}{4m a^3} - \frac{k^2}{3ma^5} \right) \,.
\label{pressurearbitrary}
\end{eqnarray}
Thus, in this limit we recover
\begin{eqnarray}
w=\frac{p_{\text{ren}}}{\rho_{\text{ren}}}\approx 0\,;
\end{eqnarray}
i.e. the matter behavior holds for any general Robertson-Walker background for 
sufficiently heavy fields.
Finally, the energy density in terms of the comoving cutoff is
\begin{eqnarray}
\rho_{\text{ren}} = \frac{1}{12\pi^2}\frac{m}{a^3}\Lambda^3_{R}=\frac{1}{12\pi^2}m\Lambda^3_{P}
\label{rren}
\end{eqnarray}
where we have only retained the dominant term and we have defined the time-dependent
physical cutoff $\Lambda_P=\Lambda_Ra^{-1}$. As it happens in other areas of physics, the cutoff value can be constrained observationally.
As we have discussed in the Introduction, the amount of energy corresponding to the vacuum of this field cannot exceed the total amount of
energy of the Universe in its different stages. In order to write the constraint on the physical cutoff $\Lambda_{P}$, it is interesting
to compare (\ref{rren}) with the total radiation energy density at a given temperature $T$:
\begin{eqnarray}
\rho_R = \frac{\pi^2}{30}\;g_*(T)\; T^4,
\end{eqnarray}
where $g_*(T)$ are the energetic effective number of relativistic degrees of freedom. Therefore, the radiation dominated epoch imposes
the following limit at a particular temperature $T$
\begin{eqnarray}
&&\Lambda_{P}\lesssim 1.46 \times 10^{-3}\, T \nonumber\\
&&\;\;\;\;\;\;\;\;\;\;\left(\frac{T}{3\, \text{eV}}\right)^{1/3}\left(\frac{m}{125\, \text{GeV}}\right)^{-1/3}\left(\frac{g_*}{3.36}\right)^{1/3}\,,
\end{eqnarray}
where we have used reference values consistent with the scalar resonance (compatible with the Higgs field) measured at the LHC: $m\simeq 125\, \text{GeV}$,
and typical values of matter radiation equality: $T\simeq 3\, \text{eV}$ and $g_* \simeq 3.36$. It is easy to understand that the strongest bound
is achieved at the end of the radiation dominated stage. At this point, the constraint coincides (approximately) with the one that determines the matter dominated universe,
\begin{eqnarray}
&&\Lambda_{P}\lesssim 0.89 \times 10^{-3}\, T \left(\frac{m}{125\, \text{GeV}}\right)^{-1/3}\,.
\end{eqnarray}
Note that if the physical cutoff saturates the above inequality, the observed dark matter content of the Universe could be explained
as the vacuum energy corresponding to a particular scalar field.

The  required physical cutoff  is a few orders of magnitude
smaller than the radiation temperature at any time.  As commented before,  such  
low momentum cutoffs appear 
naturally in the context of the holographic entropy bounds to the number of quantum states
of the gravitating system \cite{holo2,holo3}.


\section{The speed of sound of vacuum energy}


In the previous sections we have shown that a  contribution that behaves as
nonrelativistic matter
is present in the renormalized vacuum energy-momentum tensor. 
Since this matter contribution comes from the zero-point energy it 
is stable and in principle could play the role of cold dark matter.
However in
order to determine whether this contribution can actually play such a role, it is necessary to study the behavior of the density perturbations of the  vacuum energy. A viable cold dark matter fluid
should be able to allow the growth of structures, and this requires that for sub-Hubble scales
the corresponding speed of sound satisfies $c_s\ll 1$.

Let us then consider scalar perturbations around the flat Robertson-Walker background. We will
work in the longitudinal gauge, for which the perturbed metric reads
\begin{eqnarray}
\text{d}s^2 = a^2(\eta) \left\{ \left[1 + 2 \Phi(\eta,{\bf x})\right]\, \text{d}\eta^2 - \left[1 - 2\Psi(\eta,{\bf x})\right]\,\text{d}{\bf x}^2 \right\}\,.\nonumber\\
\end{eqnarray}The scalar field can also be expanded around the unperturbed solution as
\begin{eqnarray}
\phi(\eta,{\bf x}) =\phi_0(\eta,{\bf x})+\delta\phi(\eta,{\bf x})
\end{eqnarray}
where $\phi_0$ satisfies
\begin{eqnarray}
\phi_0''+2\phi_0'{\cal H}-\nabla^2\phi_0+m^2a^2\phi_0=0
\end{eqnarray}
and the total field can be shown to satisfy
up to first order in perturbations
\begin{eqnarray}
\phi''&+&(2{\cal H}-\Phi'-3\Psi')\phi'- (1+2(\Phi+\Psi))\nabla^2\phi\nonumber \\
&-&\boldsymbol{\nabla}\phi\cdot\boldsymbol{\nabla}(\Phi-\Psi)
+m^2a^2(1+2\Phi)\phi=0\,. \label{pert}
\end{eqnarray}

In order to quantize the perturbed field, we will look for a complete orthonormal set
of  solutions of the above equation. For that purpose, we will try a WKB ansatz
for the solutions in the form
\begin{eqnarray}
\phi_k(\eta,{\bf x})=f_k(\eta,{\bf x}) \,e^{i\theta_k(\eta,{\bf x})}\,,
\end{eqnarray}
where $f_k(\eta,{\bf x})$ is a slowly evolving function of $\eta$ and ${\bf x}$, whereas
$\theta_k(\eta,{\bf x})$ is a rapidly evolving  phase. Notice that such an 
approximation would work in the IR regime in which $ma$ is larger than
any other scale in the problem.

These mode solutions should be orthonormal with respect to the scalar product
\begin{eqnarray}
(\phi_p,\phi_q)&=& \\
&-&i\int_\Sigma \left[\phi_p(x)\partial_\mu \phi_q^*(x)
-(\partial_\mu \phi_p(x))\phi_q^*(x)\right]\sqrt{g_\Sigma}
\text{d}\Sigma^\mu \nonumber 
\label{scalar}
\end{eqnarray}
where $\text{d}\Sigma^\mu=n^\mu \text{d}\Sigma$  with $n^\mu$ a unit temporal vector
directed to the future and orthogonal to the $\eta=\text{const}$ hypersurface $\Sigma$, i.e.
\begin{eqnarray}
\text{d}\Sigma^\mu=\text{d}^3{\bf x} \left(\frac{1-\Phi}{a},0,0,0\right)
\end{eqnarray}
and
\begin{eqnarray}
\sqrt{g_\Sigma}=a^3(1-3\Psi)
\end{eqnarray}
to first order in perturbations.
 Thus we have
\begin{eqnarray}
(\phi_p,\phi_q)=\delta^{(3)}({\bf p}-{\bf q})
\label{normalization}
\end{eqnarray}
so that we can quantize
\begin{eqnarray}
\phi(\eta,{\bf x})=\int \text{d}^3{\bf k} \left( a_{{\bf k}}\phi_k(\eta,{\bf x})+a^{\dag}_{{\bf k}}\phi_k^*(\eta,{\bf x})\right)
\end{eqnarray}
in such a way that the corresponding creation and annihilation operators satisfy the
usual commutation relation
\begin{eqnarray}
[a_{{\bf p}},a^{\dag}_{{\bf q}}]=\delta^{(3)}({\bf p}-{\bf q})\,.
\end{eqnarray}
The positive frequency solutions of the unperturbed equation with momentum ${\bf k}$
can be written as
\begin{eqnarray}
\phi^{(0)}_k(\eta,{\bf x})=F_k(\eta) e^{(i{\bf k}\cdot{\bf x}-i\int^\eta\omega(\eta')\text{d}\eta')}
\end{eqnarray}
so that we can expand the perturbed fields as
\begin{eqnarray}
f_k(\eta,{\bf x})&=&F_k(\eta)+\delta f_k(\eta,{\bf x})\nonumber \\
\theta_k(\eta,{\bf x})&=&-\int^\eta\omega(\eta')\,\text{d}\eta'+{\bf k}\cdot {\bf x}+\delta\theta_k(\eta,{\bf x})
\end{eqnarray}
and substituting in (\ref{pert}), we get to the leading $\Od(\theta^2)$ order in the
WKB expansion
\begin{eqnarray}
-\theta'^2_k+(\boldsymbol{\nabla}\theta_k)^2(1+2(\Phi+\Psi))+m^2a^2(1+2\Phi)=0\,.\nonumber \\
\end{eqnarray}
We now expand this equation in metric perturbations so that to the lowest order
we get
\begin{eqnarray}
\omega^2=k^2+m^2a^2
\end{eqnarray}
and to the first order in perturbations
\begin{eqnarray}
2\omega\,\delta\theta'_k+2k^2(\Phi+\Psi)+2{\bf k}\cdot \boldsymbol{\nabla}\delta\theta_k
+2m^2a^2\Phi=0\,.\label{leadingpert}
\end{eqnarray}
The next term $\Od(\theta)$ of  (\ref{pert}) in the WKB expansion reads
\begin{eqnarray}
2f'_k\theta'_k&+&f_k\theta''_k+f_k\theta'_k(2{\cal H}-\Phi'-3\Psi')\nonumber \\
&-&2\boldsymbol{\nabla} f_k\cdot\boldsymbol{\nabla}\theta_k-f_k\nabla^2\theta_k=0
\end{eqnarray}
which can be expanded in turn in metric perturbations, so that to the lowest order we get
\begin{eqnarray}
-2F'_k\omega-F_k\omega'-2F_k{\cal H}\omega=0
\label{F}
\end{eqnarray}
whose solution implies that
\begin{eqnarray}
F_k(\eta)=\frac{C}{a\sqrt{2\omega}}
\label{Fsol}
\end{eqnarray}
with $C=(2\pi)^{-3/2}$ the normalization constant. To first order in metric perturbations we get
\begin{eqnarray}
&-&2\omega\delta f'_k+2F'_k\delta\theta'_k+F_k\delta\theta''_k-\omega'\delta f_k\nonumber \\
&-&2\omega{\cal H}\delta f_k+\omega F_k\Phi'+3\omega F_k\Psi'+2 F_k {\cal H}\delta\theta'_k\nonumber \\
&-&2{\bf k}\cdot \boldsymbol{\nabla}\delta f_k-F_k\nabla^2\delta\theta_k =0\,. \label{nonleadingpert}
\end{eqnarray}
In order to solve the perturbed equations (\ref{leadingpert}) and (\ref{nonleadingpert}), notice
that according to the previous discussion,  we are interested in the case in which $m^2a^2\gg k^2$.
 Thus neglecting terms $\Od(k/(ma))$ we can obtain from (\ref{leadingpert})
\begin{eqnarray}
\delta\theta'_k\simeq-ma\,\Phi\simeq -\omega\,\Phi\,.
\end{eqnarray}
Using this result and  (\ref{F}), we can rewrite  (\ref{nonleadingpert})
as
\begin{eqnarray}
\frac{1}{a\sqrt{w}}(a\sqrt{w}\delta f_k)'=\frac{3}{2}F_k\Psi'-\frac{F_k}{2\omega}\nabla^2\delta\theta_k
-\frac{{\bf k}\cdot \boldsymbol{\nabla}\delta f_k}{\omega}\,.\nonumber \\
\end{eqnarray}
However, in the limit $\omega\gg k$ we can neglect the last term, so that finally we get
using (\ref{Fsol})
\begin{eqnarray}
\delta f_k=\frac{3F_k}{2}\Psi+\frac{F_k}{2}\nabla^2\int\left(\frac{1}{\omega}\int \omega\Phi \,\text{d}\eta\right) \text{d}\eta\,.
\end{eqnarray}
We see that these solutions satisfy the normalization condition (\ref{normalization}).

On the other hand, the energy-momentum reads
\begin{eqnarray}
T^\mu_{\;\nu}=-\delta^\mu_{\;\nu}\left(\frac{1}{2}g^{\rho\sigma}\partial_\rho\phi
\partial_\sigma\phi-V(\phi)\right)+g^{\mu\rho}\partial_\rho\phi
\partial_\nu\phi\,.\nonumber \\
\end{eqnarray}
Thus to first order in perturbations we get
\begin{eqnarray}
&\rho&\,=\langle 0\vert T^{0}_{\;0}\vert 0\rangle =\\ &&\int \frac{\text{d}^3{\bf k}}{2a^2} \left((1-2\Phi)\vert{\phi'_{k}}\vert^2+(1+2\Psi)\vert\boldsymbol{\nabla}\phi_{k}\vert^2 + m^2a^2\vert\phi_{k}|^2\right)\nonumber
\end{eqnarray}
\begin{eqnarray}
&p&\,=-\langle 0\vert T^{i}_{\;i}\vert 0\rangle=\\ &&\int \frac{\text{d}^3{\bf k}}{2a^2} \left((1-2\Phi)\vert{\phi'_{k}}\vert^2
-\frac{ (1+2\Psi)}{3}\vert\boldsymbol{\nabla}\phi_{k}\vert^2 
- m^2a^2\vert\phi_{k}|^2\right)\,.\nonumber
\end{eqnarray}
Thus the first term in the energy density reads
\begin{eqnarray}
\frac{1}{2a^2}(1-2\Phi)\vert\phi_k'\vert^2=\frac{1}{2a^2}F_k^2\omega^2+F_km^2\delta f_k\,,
\label{kin}
\end{eqnarray}
the second term is negligible in the
limit $\omega\gg k$, i.e. $\vert\boldsymbol{\nabla}\phi_{k}\vert^2\ll a^{-2}\vert\phi_k'\vert^2$,
so that we can ignore it, and the last term reads
\begin{eqnarray}
\frac{1}{2}m^2\vert\phi_k\vert^2=\frac{1}{2}m^2F_k^2+F_km^2\delta f_k\,.
\label{pot}
\end{eqnarray}

Thus, adding together the contributions we see that to lowest order in perturbations we recover the results derived in previous sections,
\begin{eqnarray}
\rho^{(0)}&=& \int\frac{\text{d}^3{\bf k}}{(2\pi)^3a^4}\frac{1}{2}\omega=
 \int_0^{\Lambda_R} \frac{\text{d}{k}}{(2\pi)^2a^3} mk^2
\end{eqnarray}
whereas for the perturbation we get
\begin{eqnarray}
\delta\rho&=& \int\text{d}^3{\bf k}\,2m^2F_k\delta f_k\nonumber \\
&=&
 \int_0^{\Lambda_R} \frac{\text{d}{k}}{2\pi^2a^3} k^2m\left(\frac{3}{2}\Psi+\frac{1}{2}\nabla^2\int\left(\frac{1}{a}\int a\Phi \,\text{d}\eta\right) \text{d}\eta\right)\,,
\label{densityp}
\nonumber \\
\end{eqnarray}
and to first order in perturbations, we see that the contributions for the pressure from the
kinetic (\ref{kin}) and potential (\ref{pot}) terms cancel each other so that
\begin{eqnarray}
\delta p&=& 0\,.
\end{eqnarray}
Thus, for the corresponding speed of sound we get
\begin{eqnarray}
c_s^2=\frac{\delta p}{\delta \rho}= 0\,.
\end{eqnarray}

Notice that the $T^i_{\; j}$ components with $i\neq j$ are
\begin{eqnarray}
T^i_{\;j}=-a^{-2}(1+2\Psi)\partial_i\phi
\partial_j\phi
\end{eqnarray}
which can be neglected when compared to $T^0_{\;0}$ since as
commented before if $ma\gg k$, it is possible to neglect the spatial derivatives of the field $\phi$ as compared
to the temporal ones.
 For this reason, using Einstein equations the anisotropic stress vanishes, 
 i.e. $\Phi=\Psi$.

In the matter dominated era, for a  fluid with $c_s^2=0$, we expect 
$\Phi=\Psi=\text{const}$ both on sub-Hubble and super-Hubble scales, and
$\delta \rho/\rho\propto a$. Notice that
this is indeed what the solution of (\ref{densityp}) for $\delta \rho$ predicts since in that case
\begin{eqnarray}
\delta\rho=
 \int_0^{\Lambda_R} \frac{\text{d}{k}}{2\pi^2a^3} k^2m\left(\frac{3}{2}\Psi+\frac{1}{12}\nabla^2\Phi\eta^2\right)\propto a^{-2}
\end{eqnarray}
where we have made use of the fact  that for sub-Hubble modes $\nabla^2\Phi\,\eta^2\gg \Phi$ so that
the second term dominates.

Thus we see that not only at the background level does the vacuum energy behave as
nonrelativistic matter, but also density perturbations have negligible speed of sound
on sub-Hubble scales, which
implies the quite unexpected result  that structures could be formed out of the vacuum.
From the density perturbation and the total density we obtain the following result for
the density contrast:
\begin{eqnarray}
\frac{\delta\rho}{\rho^{(0)}}=3\Psi+\nabla^2\int\left(\frac{1}{a}\int a\Phi \,\text{d}\eta\right) \text{d}\eta
\end{eqnarray}
which is cutoff independent and valid for any scale. As a matter of fact, it agrees with the standard result for hydrodynamical
fluids \cite{Mukhanov}. From these results we see that, as in  standard cold dark matter
scenarios, 
 the growth of structures is suppressed
during the radiation dominated era and the density contrast can only start to
grow in the matter era.

\section{Conclusions}
In this work we have explored the possibility of defining the renormalized vacuum energy-momentum
tensor for massive  fields in an expanding universe by means of a constant comoving momentum cutoff. Although we have
illustrated this idea with scalar fields, the results would hold for any massive bosonic or fermionic
degree of freedom. 

We have shown that this regularization
procedure allows one to obtain a covariantly conserved renormalized energy-momentum tensor without the need of introducing noncovariant counterterms. The behavior of the vacuum energy is then shown
to depend on the  relative size of the comoving mass of field  ($am$) with respect to the cutoff. 
For large cutoffs, the UV modes dominate and the vacuum energy has the different contributions which 
have already been discussed  in the literature \cite{Maggiore}. In the case of low cutoffs (large masses) or late times, the IR modes dominate and the vacuum energy behaves as 
nonrelativistic matter. This result holds in any Robertson-Walker background and seems to 
be independent of the UV behavior. Moreover, 
 vacuum energy density perturbations in this regime are shown to have a low speed of
sound which implies that large scale structures could be seeded by vacuum energy fluctuations.

\label{sec:conclusions}



\vspace{0.5cm}
\subsection{Acknowledgments}

This work has been supported by MICINN (Spain) Project No.\ FIS2011-23000, No.\ FPA2011-27853-01 and Consolider-Ingenio MULTIDARK No.\ CSD2009-00064.
F.D.A.\ acknowledges financial support from the UAM+CSIC Campus
of International Excellence (Spain) and the kind hospitality of the
Instituto de Astrof\'isica de Canarias (IAC) while writing the manuscript.


\onecolumngrid

\appendix*


\section{Exact expressions for the energy density and pressure}


For reference, we present in this appendix the exact expressions calculated for the energy density and pressure of a massive scalar field in a de Sitter space-time.


\subsection{Energy density}


By straightforward calculation from \eqref{deSitter} and using \eqref{solution}, the energy density of a massive scalar field in a de Sitter stage is
\begin{eqnarray}
\rho_{\text{dS}} =
\frac{1}{2a^2} \int \text{d}^3{\bf k} \left(|\phi'_{k}|^2+  k^2 |\phi_{k}|^2 + m^2 a^2|\phi_{k}|^2\right)\nonumber \,
\end{eqnarray}
\begin{eqnarray}
=\,\left|\Theta_{\nu}\right|^2 \, \frac{H^4}{2} \int \frac{\text{d}^3{\bf k}}{(2\pi)^3}\, k \eta^4 \left\lbrace \left( \left| -\nu+\frac{3}{2} \right|^2 + \frac{9}{4} - \nu^2 \right) |h_{\nu-1/2}|^2 + k^2 \eta^2 \left( |h_{\nu-1/2}|^2 + |h_{\nu-3/2}|^2 \right) \right. \nonumber \\
 \left. - 2\,k\, \eta \, \text{{\bf Re}}\left[ \left( -\nu+\frac{3}{2} \right) h_{\nu-1/2}\, h^*_{\nu-3/2} \right] \right\rbrace\,,\ \ \ \
\label{energydensitydS}
\end{eqnarray}
where we have omitted the argument $-k \eta$ of the Hankel functions and the mass has been replaced using \eqref{numasa}.

In the case $m_{H}<3/2$, the last expression can be simplified to yield
\begin{eqnarray}
\rho^{m_{H}<3/2}_{\text{dS}} =  \frac{H^4}{4} \int \frac{\text{d}^3{\bf k}}{(2\pi)^3}\, k \, \eta^4  \left[ \left( \frac{9}{2} -3 \nu \right) |h_{\nu-1/2}|^2 + k^2 \eta^2 \left(|h_{\nu-1/2}|^2+|h_{\nu-3/2}|^2\right) \right. \nonumber \\
 \left. - 2\,k\, \eta \left( -\nu+\frac{3}{2} \right) \text{{\bf Re}}\,\left(h_{\nu-1/2}\,h^*_{\nu-3/2}\right) \right]\,.
\end{eqnarray}
It can easily be verified that if we put $\nu=3/2$ ($m=0$) in the above expression, we obtain the same result for the energy density of a massless field \eqref{energydensitym0dS}.

When $m_{H} > 3/2$, the energy density results
\begin{eqnarray}
\rho^{m_{H}>3/2}_{\text{dS}} = e^{-\pi \mu} \, \frac{H^4}{4} \int \frac{\text{d}^3{\bf k}}{(2\pi)^3}\, k\, \eta^4  \left\lbrace \left( \frac{9}{2} +2 \mu^2 \right) |h_{i\mu-1/2}|^2 + k^2 \eta^2 \left(|h_{i\mu-1/2}|^2+|h_{i\mu-3/2}|^2\right) \right. \nonumber \\
\left. - 2\,k\, \eta \, \text{{\bf Re}}\left[ \left( -i \mu+\frac{3}{2} \right) h_{i \mu-1/2}\,h^*_{i \mu-3/2} \right] \right\rbrace\,,\ \ \ \
\end{eqnarray}
where we have introduced $\mu=\text{{\bf Im}}(\nu)= - i \nu$. The exponential damping factor $e^{-\pi \mu}$ is not relevant since it cancels with a factor $e^{\pi \mu}$ coming from the spherical Hankel functions.
%


\subsection{Pressure}


By the same procedure, the pressure of a massive scalar field in a de Sitter stage is given by
\begin{eqnarray}
p_{\text{dS}} = \frac{1}{2a^2} \int \text{d}^3{\bf k} \left(|\phi'_{k}|^2- \frac{1}{3} k^2 |\phi_{k}|^2 - m^2a^2 |\phi_{k}|^2\right)\nonumber \,
\end{eqnarray}
\begin{eqnarray}
=\,\left|\Theta_{\nu}\right|^2 \, \frac{H^4}{2} \int \frac{\text{d}^3{\bf k}}{(2\pi)^3} \,k \, \eta^4 \left\lbrace \left( \left| -\nu+\frac{3}{2} \right|^2 - \frac{9}{4} + \nu^2 \right) |h_{\nu-1/2}|^2 + k^2 \eta^2 \left( - \frac{1}{3} |h_{\nu-1/2}|^2 + |h_{\nu-3/2}|^2 \right) \right. \nonumber \\
\left. - 2\,k\, \eta \, \text{{\bf Re}}\left[ \left( -\nu+\frac{3}{2} \right) h_{\nu-1/2}\, h^*_{\nu-3/2} \right] \right\rbrace\,,\ \ \ \
\label{pressuredS}
\end{eqnarray}
where again we have omitted the argument $-k \eta$ of the Hankel functions and the mass has been replaced using \eqref{numasa}.

When $m_{H}<3/2$, the above expression results in
\begin{eqnarray}
p^{m_{H}<3/2}_{\text{dS}} =  \frac{H^4}{4} \int \frac{\text{d}^3{\bf k}}{(2\pi)^3}\, k\, \eta^4  \left[ \nu \left(2\nu-3 \right) |h_{\nu-1/2}|^2 + k^2 \eta^2 \left(-\frac{1}{3}|h_{\nu-1/2}|^2+|h_{\nu-3/2}|^2\right) \right. \nonumber \\
 \left. - 2\,k\, \eta \left( -\nu+\frac{3}{2} \right) \text{{\bf Re}}\,\left(h_{\nu-1/2}\,h^*_{\nu-3/2}\right) \right]\,.
\end{eqnarray}
In the case $m_{H}>3/2$, the pressure is
\begin{eqnarray}
p^{m>3/2}_{\text{dS}} = e^{-\pi \mu} \, \frac{H^4}{4} \int \frac{\text{d}^3{\bf k}}{(2\pi)^3}\, k\, \eta^4  \left\lbrace k^2 \eta^2 \left(-\frac{1}{3}|h_{i\mu-1/2}|^2+|h_{i\mu-3/2}|^2\right) \right. \nonumber \\
 \left. - 2\,k\, \eta \, \text{{\bf Re}}\left[ \left( -i\mu+\frac{3}{2} \right) h_{i\mu-1/2}\,h^*_{i\mu-3/2}\right] \right\rbrace\,.
\end{eqnarray}
Again, the exponential factor cancels out when multiplied by the square modulus of the spherical Hankel functions.

\twocolumngrid

\end{document}